\documentclass[aps,pra,twocolumn,showpacs,nofootinbib]{revtex4-1}
\usepackage{graphicx,epstopdf,amsmath,amssymb,bm,color,physics}
\usepackage{color}

\newcommand{\beq}{\begin{equation}}
\newcommand{\eeq}{\end{equation}}

\allowdisplaybreaks

\begin{document}

\title{Pairing in two-dimensional Fermi gases with a coordinate-space potential}

\author{Tash Zielinski}
\affiliation{Department of Physics, University of Guelph, Guelph, Ontario N1G 2W1, Canada}
\author{Bernard Ross}
\affiliation{Department of Physics, University of Guelph, Guelph, Ontario N1G 2W1, Canada}
\author{Alexandros Gezerlis}
\affiliation{Department of Physics, University of Guelph, Guelph, Ontario N1G 2W1, Canada}

\begin{abstract}
In this work we theoretically study pairing in two-dimensional Fermi gases, a system which is 
experimentally accessible using cold atoms. We start by deriving the mean-field pairing
gap equation for a coordinate-space potential with a finite interaction range, and proceed to solve
this numerically. We find that for sufficiently short effective ranges the answer is identical 
to the zero-range one. We then use Diffusion Monte Carlo to evaluate
the total energy for many distinct particle numbers; we employ several variational parameters 
to produce a good ground-state energy and then use these results to extract
the pairing gap across a number of interaction strengths in the strongly interacting two-dimensional crossover.
Extracting the gap via the odd-even energy staggering, our microscopic results can be used as benchmarks for other theoretical approaches.
\end{abstract}


\maketitle

\section{Introduction}

The study of ultracold atomic gases has witnessed tremendous progress over the last
two decades \cite{Bloch:2008,Giorgini:2008,Levinsen:2014}. 
Historically, the interactions between different species of particles were fixed by nature; 
when the binding was strong enough to produce bosonic dimers, one was faced with
Bose-Einstein-Condensates (BEC), whereas when the interaction was weaker (but still attractive)
one dealt with Bardeen-Cooper-Schrieffer (BCS) Theory. This all changed with the development of Feschbach resonances, which use magnetic fields to tune particle interactions and allow direct experimental measurements of cold Fermi gases 
in the intermediate BEC-BCS crossover region. 

This has enabled a detailed study of the BEC-BCS crossover with direct comparison between theory and experiment, leading to the discovery of the existence of a unitary regime displaying scale invariant properties deep within the crossover region. 
The BCS theory is not expected to give quantitatively reliable results for ground-state properties in this crossover region.
Instead many-body approaches that describe the system from first principals are used. Some of the most
successful such approaches applied to three-dimensional Fermi gases
are Quantum Monte Carlo (QMC) methods, used at both zero 
and finite temperature~\cite{Carlson:2003,Astrakharchik:2004,Chang:2004,Forbes:2011,Gandolfi:2011,Forbes:2012,Carlson:2012,Pilati:2014,
Schonenberg2:2017,Dawkins:2017}.

Reduced dimensionality -- specifically 2D systems -- are a rich area which has been at the forefront
of research more recently, as they display distinct and unique properties compared to that of the 3D gas systems.
Experimentally, cold Fermi gases are produced in quasi-2D pancake-shape gas clouds~\cite{Gunter:2005,Martiyanov:2010,Frohlich:2011,Feld:2011,Orel:2011,Dyke:2011,Sommer:2012,Makhalov:2014,Boettcher:2015,Ong:2015,Murthy:2015,Ries:2015,Fenech:2016,Boettcher:2016,Martiyanov:2016,Cheng:2016,Luciuk:2017,Toniolo:2017}. 
Both preceding and following the experimental breakthroughs, a number of theoretical approaches
have been brought to bear on the subject of lower-dimensional Fermi gases~\cite{Nishida:2008,Liu:2010,Valiente:2011,Bertaina:2011,Werner:2012,Bauer:2014,Shi:2015,Anderson:2015,He:2015,Rammelmuller:2015,He:2016,Galea:2016,Klawunn:2016,Galea:2017,Vitali:2017,Schonenberg:2017,Drut:2018,jeszenszki:2019}.

In addition to being experimentally accessible and therefore interesting in their own right, cold Fermi atomic gases can also
help shed light on the physics of superfluidity in neutron-star crusts~\cite{Gezerlis:2008,Gandolfi:2015,Buraczynski:2016}. 
In a sense, neutron matter is an attractive Fermi gas where one needs to take the effective range into account,
in addition to the scattering length. While the neutron-neutron interaction is fixed by nature,
one can envision cold-atomic experiments which probe the regime of relevance to neutron stars~\cite{Horikoshi:2017}.
More generally, two-dimensional cold gases can also help us understand
nuclear ``pasta'' phases, where nucleons end up populating highly deformed quasi-1D or quasi-2D systems. 

In earlier works, we have employed Diffusion Monte Carlo~\cite{Galea:2016,Galea:2017} (DMC) to evaluate
two-dimensional cold-gas properties such as the equation of state and the pairing gap. These involved, first,
the use of a finite-range two-particle interaction and, second, the use of only $N=26$ particles in a periodic area.
In the present paper, we investigate the effects of both of these choices in more detail. Specifically, we start from
investigating the effect of a finite effective range in the context of another many-body approach, namely mean-field
BCS theory; generally speaking, this allows us to use a simpler theory to study effects which would be much harder
to tackle in the context of DMC. Then, we turn to our new DMC results for the energy and the pairing gap, which 
result from using larger-$N$ values; as part of this process, we try to systematize our understanding of the finite-size
effects along the 2D crossover.

\section{Two-body binding energy}
\label{sec:twobody}

For the two-particle problem in two dimensions~\cite{Miyake:1983,Randeria:1989},
the binding energy is: 
\begin{equation}
\label{eq:bindingEnergy}
e_b^0 = -\frac{4\hbar^2}{m a^2 e^{2\gamma}}
\end{equation}
where $a$ is the 2D scattering length, $m$ is the particle mass, and $\gamma\approx0.577215$ also known as the Euler number. 
The Fermi energy is:
\begin{equation}
E_F = \frac{\hbar^2 k_F^2}{2m}
\label{eq:bindingEnergy2}
\end{equation}
where $k_F$ is the Fermi wave number. In two dimensions the free-gas
energy per particle is simply $E_{fg} = E_F/2$.

The expression in Eq.~(\ref{eq:bindingEnergy}) involves the scattering length, $a$,
but not the effective range, $r_e$; this is a zero-effective-range result, i.e., it is 
valid for $k_F r_e \rightarrow 0$, where $r_e$ is the effective range
of the two-particle interaction. 
On the other hand, many Quantum Monte Carlo calculations are carried out using coordinate-space potentials
of finite (but tiny) range. A popular interaction is of the (modified) P\"oschl-Teller form:
\begin{equation}
\label{eq:potential}
V(r)=\frac{-v_0 \hbar^2}{m_r}\frac{\nu^2}{\cosh^2(\nu r)}
\end{equation}
where $v_0$ and $\nu$ are parameters which we tune to match the desired scattering length $a$ and effective
range $r_e$, and $m_r$ is the reduced mass. 
As is standard for this problem, we quantify the \emph{interaction strength} by defining:
\begin{equation}
\label{eq:interactionStrength}
\eta=\ln{(k_F a)}.
\end{equation}
While the $k_F$ which appears here and in Eq.~(\ref{eq:bindingEnergy2}) is a many-body quantity
(which therefore does not arise at the two-body level), we find it convenient to employ
it to produce
dimensionless quantities throughout.

\begin{figure}[t]
\includegraphics[width=\columnwidth]{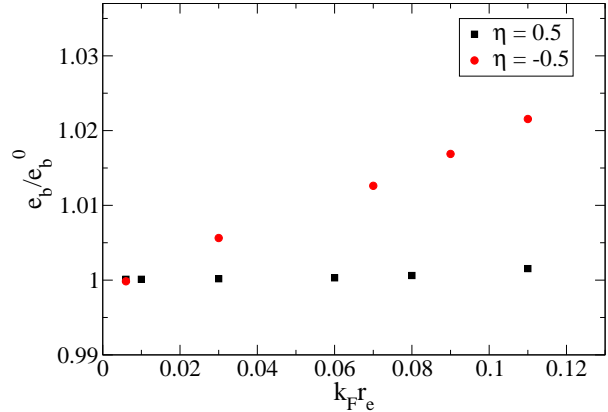}
\caption{Two-body binding energy vs effective range. 
To make the values shown on each axis dimensionless, we are forming the product 
$k_F r_e$, employing the Fermi wave number, and the ratio $e_b/e_b^0$,
employing the zero-range binding energy from Eq.~(\ref{eq:bindingEnergy}).
}
\label{fig:twobody}
\end{figure}

As further discussed in Ref.~\cite{Galea:2017}, in two dimensions 
a partial-wave expansion of the time-independent Schr\"odinger equation
leads to 
\begin{equation}\label{eq:2dse}
-\frac{\partial^2 u_0(r)}{\partial r^2} = u_0(r) \bigg[\frac{2m_r}{\hbar^2} [e_b - V(r)] + \frac{1}{4r^2}\bigg]\,.
\end{equation} 
for the $l=0$ partial wave. Here 
$u_0(r)=\sqrt{r}R_0(r)$ and $R_0(r)$ is (proportional to) the wave function. 
We have numerically solved Eq.~(\ref{eq:2dse}) to find the finite-effective-range
two-body binding energy, $e_b$, for the two cases of $\eta=-0.5$ and $\eta=0.5$.
The result is shown in Fig.~\ref{fig:twobody}:
we see that for $\eta=0.5$ the range is largely
irrelevant, since we basically get the same binding energy for all effective ranges up
to $k_F r_e \approx 0.1$. The case of $\eta=-0.5$ is different, in that the binding
energy changes by a few percent for such effective ranges. For both $\eta$ values,
the figure clearly shows that the effective range we employ in our DMC calculations,
$k_F r_e \approx 0.006$, is short enough that we don't have to worry about 
finite-range effects. As a result, in what follows we will be employing
$e_b^0$ and $e_b$ interchangeably, under the assumption that the effective range
is short enough.

\section{Mean-Field BCS Theory}
\label{sec:mean}

Bardeen-Cooper-Schrieffer theory for the two-dimensional case was developed before the recent
explosion of activity in low-dimensional cold-atom systems~\cite{Miyake:1983,Randeria:1989}. For $s$-wave pairing,
which is our focus in this paper, the problem is analytically solvable, leading to simple expressions for the pairing
gap and chemical potential. These were:
\begin{equation}
\label{eq:Randeria}
\Delta_{\rm gap} = 
    \begin{cases}
      \sqrt{2E_F|e_b|},~~~~~~~\mu>0\\
      E_F + |e_b|/2,~~~~\mu<0
     \end{cases}
\end{equation}
and
\begin{equation}
\mu = E_F + e_b/2
\end{equation}
respectively.  
Crucially, the derivation in Ref.~\cite{Randeria:1989} involved the two-body $T$-matrix, in an attempt to avoid
issues arising from the use of a coordinate-space potential which involves a hard-core repulsion.
The expressions above are valid in the limit of $k_Fr_e \rightarrow 0$.  

In our Quantum Monte Carlo calculations we employ Eq.~(\ref{eq:potential}); 
this is a purely attractive potential, so it does not give rise
to problems in momentum space.
Motivated by such studies, in this work we first derive the BCS gap equation in 2D
for a finite-range potential; to our knowledge, this derivation and the resulting expression have not been published before.
We then numerically solve the gap equation, comparing to the $k_Fr_e \rightarrow 0$ solution, thereby testing if the effective
range to be used in later sections (on DMC) is sufficiently small.

The BCS gap equation has the form:
\begin{equation}
\label{eq:gapsum}
\Delta({\bf k})=-\sum_{\bf k'} \bra{\bf k}V\ket{\bf k'} \frac{\Delta({\bf k'})}{2E({\bf k'})}
\end{equation}
Where $\Delta(\mathbf{k})$ is the pairing gap function, $V$ is the potential energy, and $E(k)$ is the quasiparticle energy, given by the relationship:
\begin{equation}
\label{eq:quasi}
E({\bf k})=\sqrt{\left(\frac{\hbar^2\abs{\bf k}^2}{2m}-\mu\right)^2+\Delta^2({\bf k})}
\end{equation}
We are here interested only in qualitative features resulting from a small but finite effective range;
thus, we are employing a free single-particle spectrum, as shown in Eq.~(\ref{eq:quasi});
a more complete study would also involve normal-state interactions.
In these equations, the gap function and potential matrix element are given in momentum space. We would like to see
how that relates to the coordinate-space potential, which could be of the form of Eq.~(\ref{eq:potential}).
We have:
\begin{equation}
\label{eq:vMatDer}
\begin{aligned}[b]
\bra{\bf k}V\ket{\bf k'} &= \int d{\bf r} \int d{\bf r'} \braket{\bf k}{\bf r}\bra{\bf r}V\ket{\bf r'}\braket{\bf r'}{\bf k'} \\
& = \frac{1}{L^2} \int d{\bf r} \int d{\bf r'} e^{i{\bf k \cdotp r}} \bra{\bf r}V\ket{\bf r'} e^{-i{\bf k' \cdotp r'}} \\
& = \frac{1}{L^2} \int d{\bf r} \int d{\bf r'} \sum_{n,m} i^n J_n(kr) e^{in(\theta_k-\theta_r)} V(r) \\
&\times \delta({\bf r}-{\bf r'}) (-i)^m J_m(k'r') e^{-im(\theta_{k'}-\theta_{r'})} \\
& = \frac{1}{L^2} \sum_{n,m} i^n(-i)^m e^{in\theta_k}e^{-im\theta_{k'}} \\
&\times \int_0^\infty dr r J_n(kr)V(r)J_m(k'r) \\
&\times \int_0^{2\pi} d\theta_r e^{i(m-n)\theta_r} \\
& = \frac{2\pi}{L^2}  \sum_{n=-\infty}^\infty e^{in(\theta_k-\theta_{k'})} V_n(k,k').
\end{aligned}
\end{equation}
In the first equality we introduced two resolutions of the identity. In the second equality
we plugged in the appropriate plane waves, normalized to fit inside a square of area $L^2$.
In the third equality we employed the locality of the interaction and 
used (twice) the two-dimensional expansion of a plane wave:
\begin{equation}
\label{eq:pwExp}
\begin{aligned}[b]
e^{i{\bf k \cdotp r}} = \sum_{n=-\infty}^\infty i^n J_n(kr) e^{in\theta}
\end{aligned}
\end{equation} 
where $J_n$ is the Bessel function of order $n$, and $\theta$ is the angle between the $\mathbf{k}$ and $\mathbf{r}$ vectors, or $\theta=\theta_k-\theta_r$. In the fourth equality we merely rearranged terms and in the fifth equality we 
carried out the integral over $\theta_r$ and used the result to eliminate one of the sums;
we also took the opportunity to introduce $V_n(k,k')$ for the Bessel-transformed potential.
The result is analogous to a Fourier transform of $V(r)$, in that when the potential in real space is narrow and deep, the potential in $k$-space is wide and shallow, and vice versa. This is important because in what follows we
will be interested precisely in extremely deep, narrow real-space potentials, making $V(k, k')$ very wide. 

We now carry out an analogous expansion for the gap function:
\begin{equation}
\label{eq:gapfourier}
\Delta({\bf k}) \equiv \sum_{n=-\infty}^\infty \Delta_n(k)e^{in\theta_k}
\end{equation}
Inserting this, as well as the equation for potential matrix element, Eq.~(\ref{eq:vMatDer}), the right-hand side
of the gap equation in  
Eq.~(\ref{eq:gapsum}) turns into:
\begin{equation}
\begin{aligned}
&-\frac{2\pi}{L^2} \sum_{\bf k'} \sum_{n,m} e^{im\theta_{k'}}e^{in(\theta_k-\theta_{k'})} V_n(k,k') \frac{\Delta_m(k')}{2E(k')} \\
&= -\frac{2\pi}{(2\pi)^2} \int d{\bf k'} \sum_{n,m} e^{i(m-n)\theta_{k'}}  e^{in\theta_k} V_n(k,k') \frac{\Delta_m(k')}{2E(k')} \\
&= -\frac{1}{2\pi} \sum_{n,m} e^{in\theta_k} \int_0^\infty dk'k' V_n(k, k') \frac{\Delta_m(k')}{2E(k')} \\
&\times \int_0^{2\pi} d\theta_{k'} e^{i(m-n)\theta_{k'}} \\
&= -\sum_n e^{in\theta_k} \int_0^\infty dk'k' V_n(k,k') \frac{\Delta_n(k')}{2E(k')} \\
\end{aligned}
\end{equation}
In the first equality, we cancelled the $L^2$ when converting from sum to integral. 
In the second equality we wrote out the integral over $\mathbf{k}'$ and in the third
eqaulity
we used the Kronecker delta to eliminate one of the sums.

If we now recall that the left-hand side of the BCS gap equation of Eq.~(\ref{eq:gapsum}) will
also take the form Eq.~(\ref{eq:gapfourier}), we can multiply both sides by $e^{-il\theta_k}$ and integrate, to get:
\begin{equation}
\begin{aligned}
\Delta_l(k) = -\int_0^\infty dk'k' V_l(k,k') \frac{\Delta_l(k')}{2E(k')}
\end{aligned}
\label{eq:gapIntDer}
\end{equation}
In what follows, we will be interested in taking $l=0$, corresponding to the $s$-wave problem. Thus, we
have arrived at the two-dimensional BCS gap equation in the thermodynamic limit. 
In practice, when carrying out the integral over $k'$ one must be careful to go up to sufficiently
high momenta where $V_0$ has died off.

In addition to the BCS gap equation, one must also self-consistently solve the constant-average-density
equation:
\begin{equation}
\label{eq:density}
\begin{aligned}
\rho = \int_0^\infty dkk \left( 1 - \frac{\frac{\hbar^2k^2}{2m}-\mu}{E(k)}\right)
\end{aligned}
\end{equation}
In practice, we solve Eq.~(\ref{eq:gapIntDer}) together with Eq.~(\ref{eq:density})
iteratively: this is fixed-point iteration for an entire function. Given that we are dealing with 
a purely attractive potential, the process converges; as part of this, we tune the chemical 
potential $\mu$ to get the desired density $\rho = k_F^2/2\pi$.

\begin{figure}
\includegraphics[width=\columnwidth]{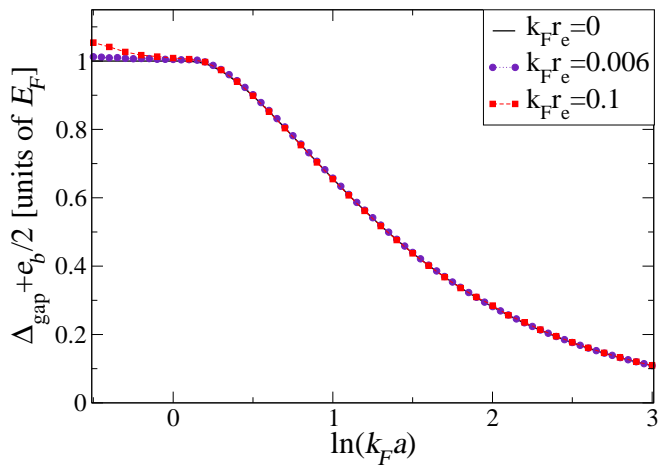}
\caption{Mean-field BCS results versus interaction strength, for different dimensionless effective ranges ($k_F r_e$). As the effective range goes towards zero, the curves more closely resemble the analytical solution of Eq.~(\ref{eq:Randeria}).}
\label{fig:meanField}
\end{figure}

Results of our mean-field calculations are given in Fig.~\ref{fig:meanField} where the solid line of $k_Fr_e=0$ is the analytical solution from Eq.~(\ref{eq:Randeria}). 
The other two curves are new results, which match very closely with the analytical solution for $k_Fr_e=0.006$, and even for the larger $k_Fr_e=0.1$ for positive interaction strengths. These results inticate that the choices of $k_Fr_e$'s used in our DMC calculations (in Refs.~\cite{Galea:2016,Galea:2017} and below) are small enough to get meaningful results. This is fully consistent with what we discovered at the two-body level 
in Fig.~\ref{fig:twobody}.

\section{Quantum Monte Carlo}
\label{sec:dmc}

The problem we are faced with is how to tackle the following two-component Hamiltonian in two dimensions:
\begin{gather}
\hat{H} = \frac{-\hbar^2}{2m} \Bigg[\sum_{i=1}^{N_\uparrow}\nabla_i^2
+\sum_{j'=1}^{N_\downarrow}\nabla_{j'}^2\Bigg]
+\sum_{i,j'}V(r_{ij'})\,,
\end{gather}
where $m$ is the particle mass and the particle number is $N_\uparrow + N_\downarrow = N$.  
The $V(r_{ij'})$ is taken to be of the modified P\"oschl-Teller form, Eq.~(\ref{eq:potential}). 
In what follows, we work in a periodic simulation area, as is standard in continuum Quantum Monte Carlo
approaches. Specifically, we work with $N$ particles in a square of length $L$; 
allowed momentum states are governed by the equation:
\begin{equation}
\label{eq:kstate}
{\bf k_n} = \frac{2\pi}{L}(n_x, n_y)
\end{equation}
where $n_x$ and $n_y$ are integers. 

\begin{figure}[b]
\includegraphics[width=\columnwidth]{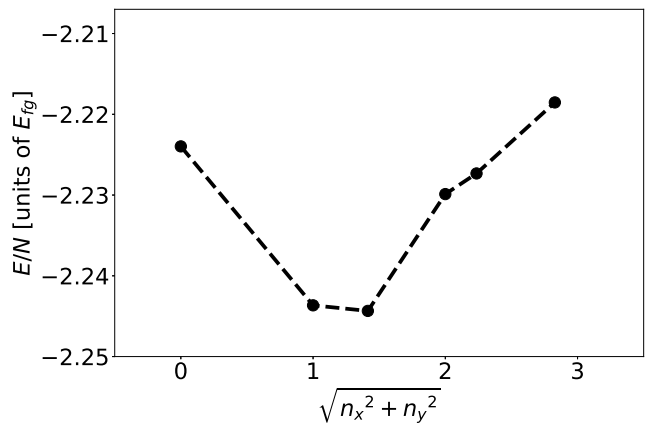}
\caption[DMC Energy with respect to chosen k-state]{DMC energy-per-particle for $N=43$ particles at
$\eta=0$ versus the quantum numbers at which the 43rd particle can be placed. A clean minimum is observed at the $(1,1)$ momentum state, see Eq.~(\ref{eq:kstate}).}
\label{fig:k_finite}
\end{figure}

The first step in our study is always a Variational Monte Carlo (VMC) calculation: this provides
an upper bound on the ground-state energy. We use $\textbf{R}$ will denote a \emph{walker} which represents the coordinates of all of the particles in the simulation box, and is $2N$ dimensional. Using a trial wave function, $\Psi_T (\textbf{R})$, we determine the variational energy, $E_v$, which is the upper bound on the ground-state energy $E_0$:
\begin{equation}
\label{eq:VMC}
E_v = \frac{\int \Psi_T^*(\textbf{R}) \hat{H} \Psi_T (\textbf{R}) d \textbf{R}}  {\int \Psi_T^*(\textbf{R}) \Psi_T  (\textbf{R}) d \textbf{R}} \ge E_0
\end{equation}
This is nothing other than the Rayleigh-Ritz principle applied to a many-particle system.
For a ``good'' wave function, this will provide a decent estimate of the ground-state energy of an 
interacting system. The many-dimensional integrals involved are carried out stochastically,
using a large number of walkers.

The question of what constitutes a good wave function naturally arises at this point.
Following Refs.~\cite{Galea:2016,Galea:2017}, we employ a 
Jastrow-BCS trial wave function:
\begin{equation}
\begin{split}
\Psi_T({\bf R}) = \prod_{ij'}f_{J}(r_{ij'}) \, \Phi_{\rm BCS} \,,~~~~~ \\
\label{eq:BCS}
\Phi_{\rm BCS} = {{\cal A}} [\phi({\bf r}_{11'}) \phi({\bf r}_{22'}) ... \phi({\bf r}_{N_\uparrow N'_\downarrow})]\,.
\end{split}
\end{equation}
where $f_{J}(r)$ is a nodeless Jastrow term, 
$\cal A$ is an antisymmetrizing operator, and we took $N_\uparrow = N_\downarrow$.
We express the pairing function $\phi({\bf r})$ as a sum of 10 plane wave terms, each multiplied
with an unknown parameter, and we capture higher-momentum contributions by a spherically
symmetric function. Then, we employ VMC to determine the best possible set of these 10 parameter values;
crucially, this is a process which we repeat for each new particle number $N$.
As a technical aside we note that in two-dimensional physics, where the $k_F a$ spans multiple orders
of magnitude, the Jastrow term is implemented using a table with substantially more points than in the 
three-dimensional problem.

\begin{figure}[b]
\includegraphics[width=\columnwidth]{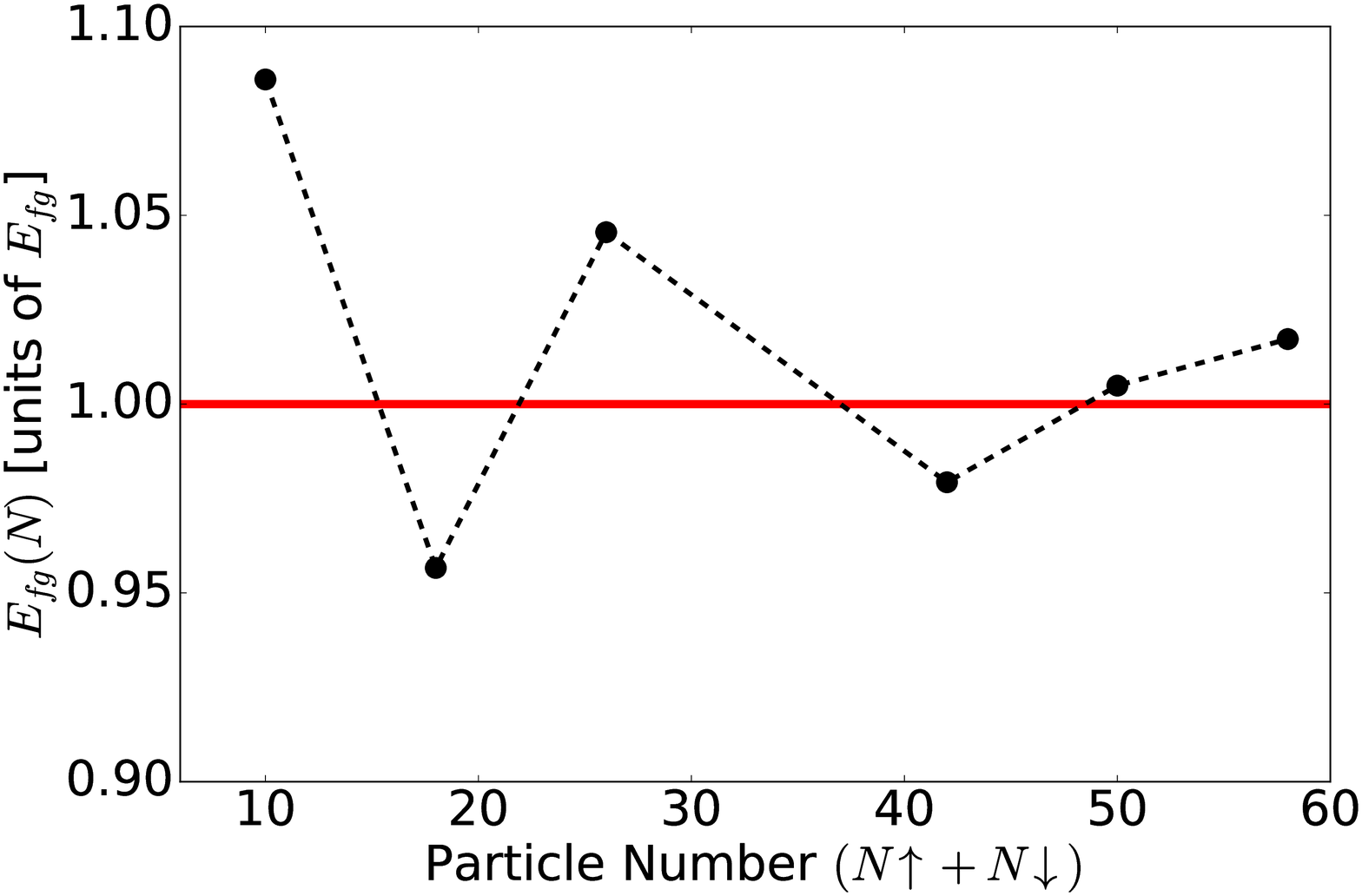}
\caption[Closed shell free gas finite size errors]{Non-interacting energy per particle as a function of
the particle number (always one unit away from a closed shell). The horizontal line is the value at the thermodynamic limit.}
\label{fig:finite_err}
\end{figure}

The next step in our calculations is to use Diffusion Monte Carlo (DMC):
this also approximates the ground-state energy using a trial wave function, but is superior
in that it propagates in imaginary time to reduce the contribution of the parts of the trial wave function that are not the ground state (i.e., excited states). This is done by solving the Schr\"{o}dinger equation in imaginary time.
Schematically: 
\begin{equation}
\Psi(\tau \rightarrow \infty) =
 \lim_{\tau \rightarrow \infty} e^{-({\cal{H}}-E_T) \tau} \Psi_T~\nonumber \rightarrow \alpha_0 e^{-(E_0-E_T) \tau} \Psi_0~\nonumber
\end{equation}
where $E_T$ is the so-called ``trial energy'' which helps us ensure the walker number doesn't get out of hand.
In practice, we propagate up to sufficiently large imaginary time, such that there is no longer a decay going on.

Apart from technical practicalities like the trial energy and the Jastrow term, the only other consideration 
is the fermion sign problem. This arises due to the presence of both positive and negative nodal pockets
in the complicated many-particle wave function. As is standard in zero-temperature continuum QMC algorithms,
we employ the fixed-node approximation, which does not allow walkers to change sign in the wave function.
Assuming the trial wave function has the appropriate physical content (e.g., pairing properties for a superfluid system), the effect of the fixed-node approximation ends up being tiny. Fortunately, both our VMC and
our DMC results obey a variational property, meaning that one can keep trying to improve the answers by
using a better wave function.

Even so, it's worth highlighting
that one of our goals in this work is to calculate the pairing gap via the odd-even staggering in the energy:
\begin{equation}
\label{eq:gap_eqn}
\Delta(N)=  E(N+1)-\frac{1}{2} \left [ E(N+2)+E(N) \right ],
\end{equation}
Here $E(N)$ is the total energy of a closed shell $N=10, 18, 26, 42, 50, 58$, $E(N+1)$ is the total energy of the system of a closed shell with one extra particle that does not have a partner particle to pair with, and $E(N+2)$ is one more particle added to the system which again causes all particles to be paired.
We have explored a couple higher closed shells, but the statistical errors were so large that it became
impossible to trust the energy 
differences; note that QMC methods employ determinants, so the computational cost typically scales
as $N^3$. That means that a calculation for $N=58$ (the largest particle number employed here) is roughly 11 times more demanding than one 
for $N=26$ (which was the particle number used in earlier works).
It's worth emphasizing that, while a variational property applies to the total energy, it does not apply to energy differences such as
in Eq.~(\ref{eq:gap_eqn}). That being said, similar QMC predictions for the problem of the three-dimensional
pairing gap \cite{Gezerlis:2008} turned out to be verified experimentally~\cite{Horikoshi:2017}.

\begin{figure}[t]
\includegraphics[width=\columnwidth]{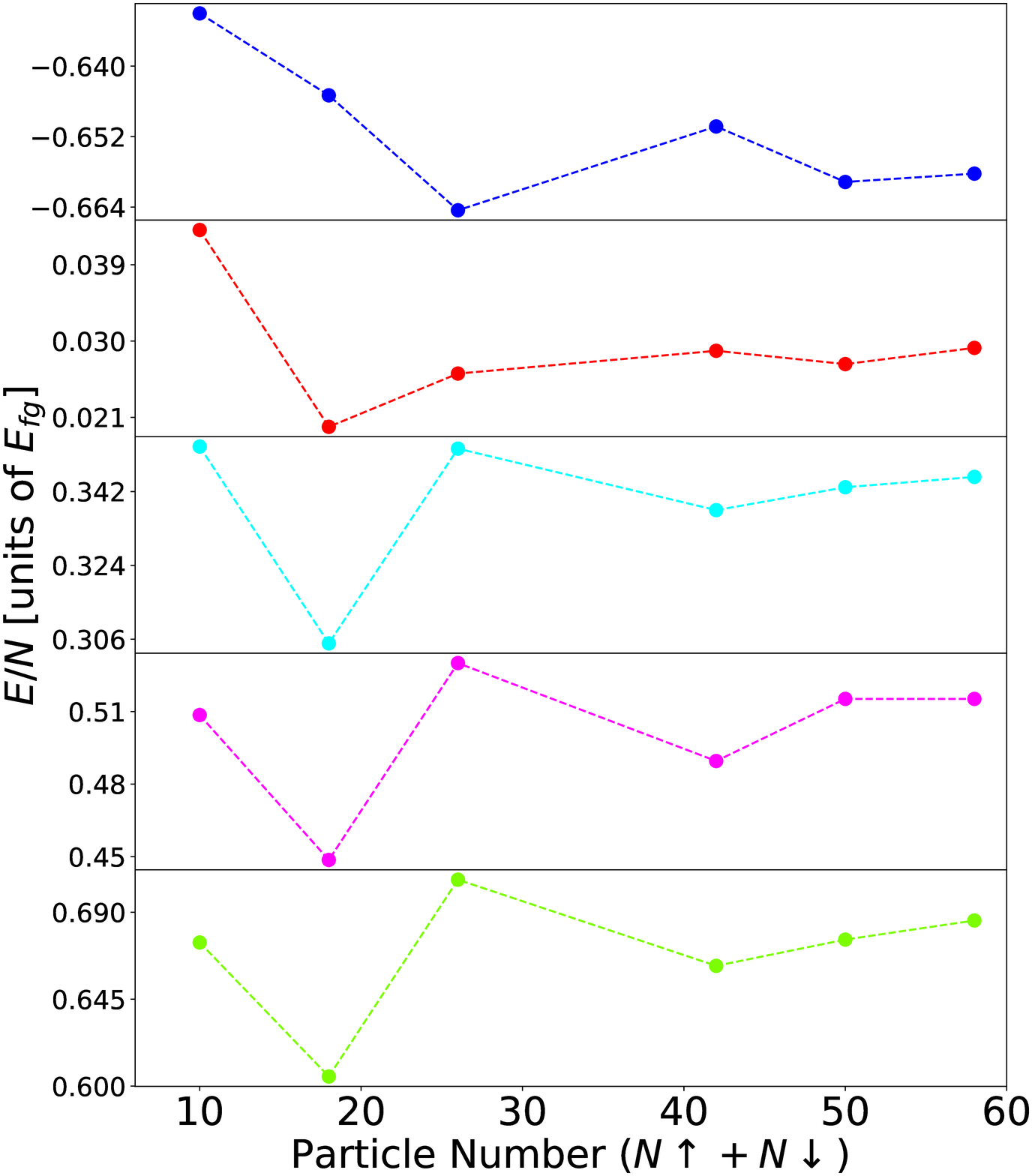}
\caption[Finite size effects energy per particle via DMC]{Exploration of the finite-size effects on the DMC energy
per particle. From top to bottom, the panels correspond to $\eta = 0.5, 1.0, 1.5, 2.0, 3.0$.}
\label{fig:all_E}
\end{figure}

With the systematic errors under control, the only things left to investigate are the finite-size effects
(i.e., the dependence on $N$) as well as the placement of the $(N+1)$-th particle. We discuss
the former in detail in the following section; as for the latter, we note that in Eq.~(\ref{eq:gap_eqn})
one must determine the pairing gap as the minimum. For fermions in a non-interacting gas, or in a normal gas,
the placement of the next particle is trivial: it goes onto the next available momentum state. 
For the present case, where pairing correlations play a major role, the choice of momentum states is not
quite so intuitive. This is illustrated in Fig.~\ref{fig:k_finite}, in which the interacting system's optimal $k$-state is not simply the next unoccupied shell. This is analogous to what we saw in Eq.~(\ref{eq:quasi}). The VMC values can be quite close and within error so we chose to include a DMC run in the determination of the $k$-states. (Incidentally, we have encountered one case
where the 27-particle energies of Ref.~\cite{Galea:2016} gave different answers in VMC and DMC; 
we have corrected this here, always employing the lowest-possible values.)

\section{Results}
\label{sec:interacting}

We start our discussion of the finite-size effects by looking at the energy per particle in the non-interacting
gas, see Fig.~\ref{fig:finite_err}.  The red horizontal line is the thermodynamic limit value (infinite volume,
infinite particle number, constant density). We see that for small particle numbers there are large
deviations from the thermodynamic-limit answer, up to 5\% or more, but these go away as the particle
number is increased. Of course, the question arises whether one should expect the interacting system
finite-size effects to match those of the non-interacting gas.

The results of our DMC calculations within the strong coupling crossover for various interaction strengths in descending order from $\eta=0.5$ to $\eta=3.0$ are shown in Fig.~\ref{fig:all_E}. Looking at the general shape of the DMC energies, the bottom three panels follow roughly the same trend where the $N=10$ is large then a local minimum occurs at $N=18$ and a local maximum at $N=26$ which proceeds to have a slight dip at $N=42$ then small increase to $N=58$; this trend closely follows the shape of the free gas plot of Fig.~\ref{fig:finite_err}. These results imply that the pairing correlations on the BCS side of the crossover region ($\eta$ greater than or equal to 1.5) have a small effect on the energy dependence on particle number. This suggests that even for
other observables one could get guidance from the non-interacting gas behavior. 

\begin{figure}[t]
\includegraphics[width=\columnwidth]{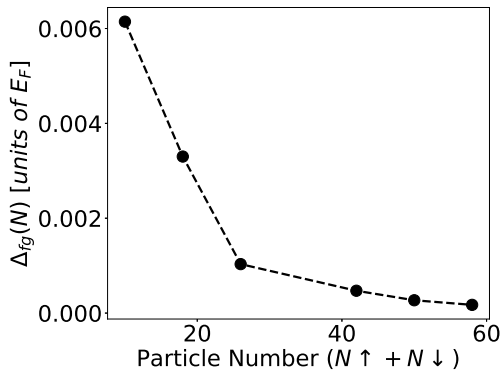}
\caption[Applying the gap perscription to the free gas]{The pairing gap of Eq.~(\ref{eq:gap_eqn}) applied
to the non-interacting gas problem. This is zero in the large-system limit; the finite values for smaller 
systems are finite-size artifacts.}
\label{fig:fg_gaps}
\end{figure}

Turning to the stronger interaction strengths (the top two panels in Fig.~\ref{fig:all_E}), 
we find fewer similarities with the non-interacting gas trend. 
Going from $N=10$ to $N=18$ there is a drop, but the commonalities largely end there. 
The panel that is the most different from Fig.~\ref{fig:finite_err} is that corresponding to 
$\eta=0.5$ , where $N=18$ does not display a local minimum. 
Overall,
we can say that the trend (as a function of $N$) is ``flatter'' for the top two panels, i.e., the results
for $N=50$ and $N=58$ are very close to each other; this is a result of stronger pairing correlations.
It's crucial to note that for small $\eta$ the finite-size effects are less significant:
for $\eta=0.5$ the fluctuation from the largest to the smallest value is roughly 5\%, to be compared
with the corresponding change of approximately 15\% in the non-interacting gas. In short,
for even smaller $\eta$ (say, $0$ or $-1$) the finite-size effects become irrelevant.

\begin{figure}[t]
\includegraphics[width=\columnwidth]{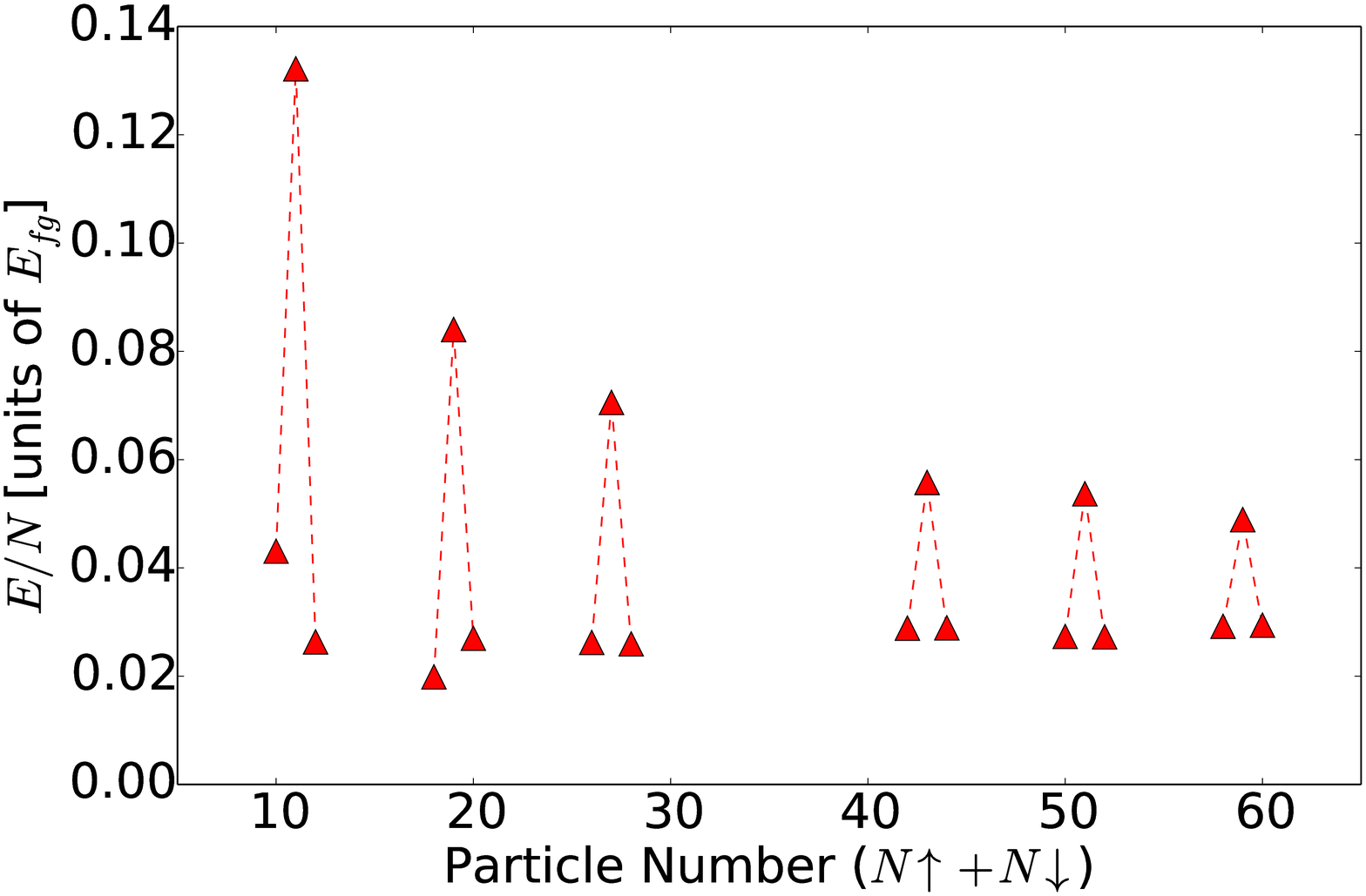}
\caption[All E/N vs N for $\eta=1.0$ ]{DMC energies at $\eta=1.0$ for the closed-shell values $N = 10, 18, 26, 42, 50, 58$ and $(N+1)$, $(N+2)$ in each case. Dashed lines connect each
triple.}
\label{fig:all_E1}
\end{figure}

Since we will be computing the pairing gap from the odd-even staggering of Eq.~(\ref{eq:gap_eqn}),
we now spend some time discussing the energies of systems with $N$, $N+1$, and $N+2$ particles.
Similarly to what we did in Fig.~\ref{fig:finite_err}, we first go over the finite-size effects in the non-interacting
gas, to possibly some guidance about how to interpret our DMC results.
In Fig.~\ref{fig:fg_gaps} we show the result of applying the odd-even staggering formula
to the non-interacting gas. Since there is no pairing in the absence of interactions, one would expect
the pairing gap to be zero. However, this figure clearly shows that at lower paticle numbers the $\Delta_{fg}$ is largest and decreases as $N$ increases. This tells us that from the particle numbers we are employing
 $N=50$ and $N=58$ should have the least amount of error. Of course, this fact should be combined
 with our earlier finding, namely that the finite-size effects of the interacting gas are insignificant for small
 $\eta$ (but track those of the non-interacting gas for large $\eta$).

\begin{figure}[t]
\includegraphics[width=\columnwidth]{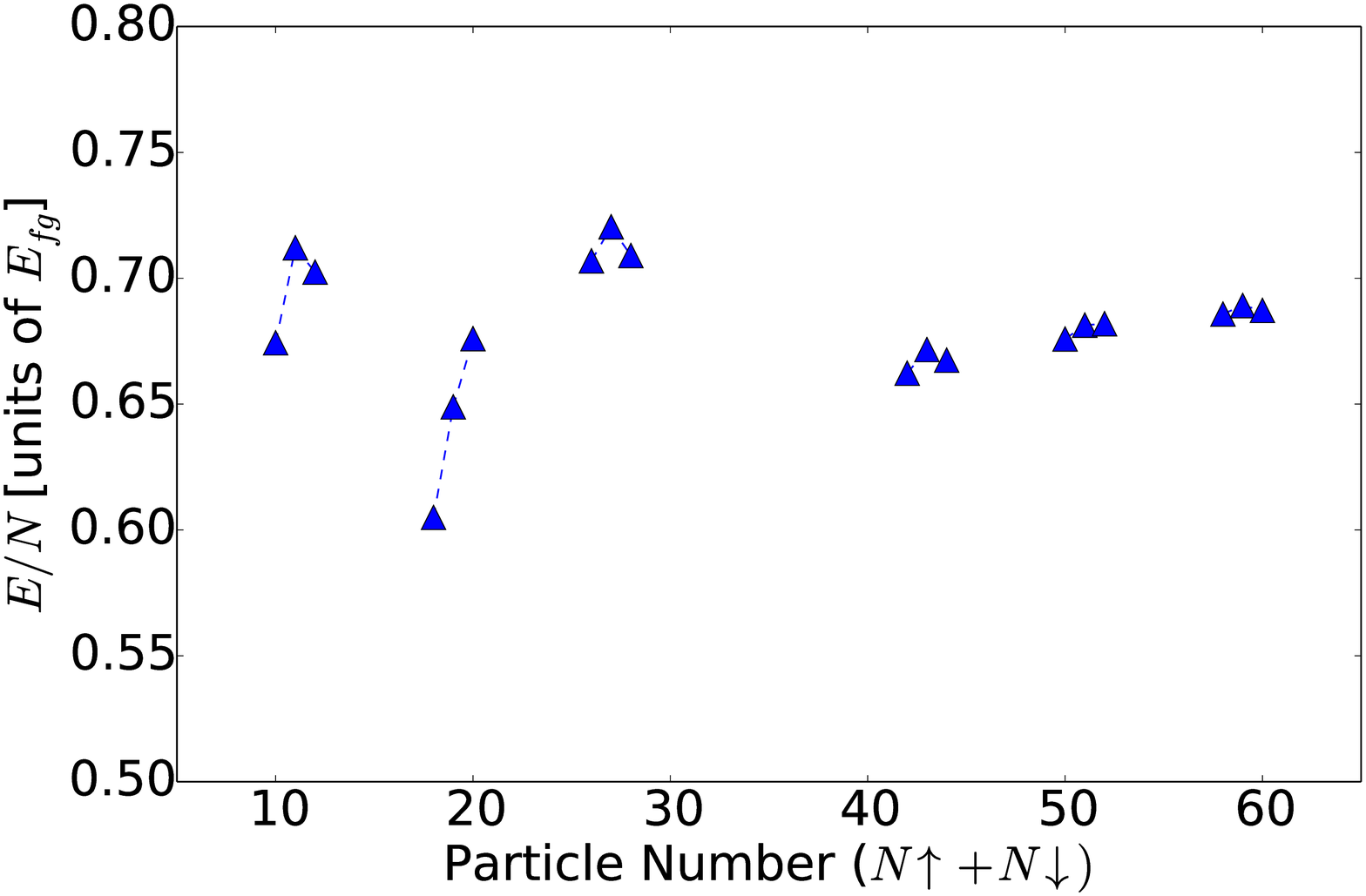}
\caption[All E/N vs N for $\eta=3.0$ ]{DMC energies at $\eta=3.0$ for the closed-shell values $N = 10, 18, 26, 42, 50, 58$ and $(N+1)$, $(N+2)$ in each case. Dashed lines connect each
triple.
Comparing to Fig.~\ref{fig:all_E1} we see quite different
behavior of each adjacent set of points.}
\label{fig:all_E3}
\end{figure}

We now turn to the energies of strongly interacting systems with $N$, $N+1$, and $N+2$ particles; 
we show results for $\eta = 1$ and $\eta = 3$ in Fig.~\ref{fig:all_E1} and Fig.~\ref{fig:all_E3}, respectively
(statistical errors are smaller than the symbols shown). 
In both cases, we observe the same general pattern: the three adjacent-in-$N$ values are quite far apart
from each other for small $N$, but then get closer as $N$ is increased, leading to the characteristic 
odd-even staggering pattern. One should not confuse this trend with the conclusion that the gap decreases as the $N$ increases, since the even partners are not placed symmetrically for small-$N$ values. 
This is not surprising given what we saw in Fig.~\ref{fig:finite_err}.
Looking back at the odd-even staggering prescription of Eq.~(\ref{eq:gap_eqn}),
it is important to realize that the energies plotted are per particle, so these need to be multiplied
by the appropriate particle numbers before the difference is taken. 
Another similarity between the two plots is that the results around
$N=50$ and $N=58$ behave in roughly the same way.
It's also worth discussing the differences between Fig.~\ref{fig:all_E1} and Fig.~\ref{fig:all_E3}.
In Fig.~\ref{fig:all_E1} the result for $N+1$ is always higher than its neighbors, a fact which is 
not always true in Fig.~\ref{fig:all_E3}.

\begin{figure}[b]
\includegraphics[width=\columnwidth]{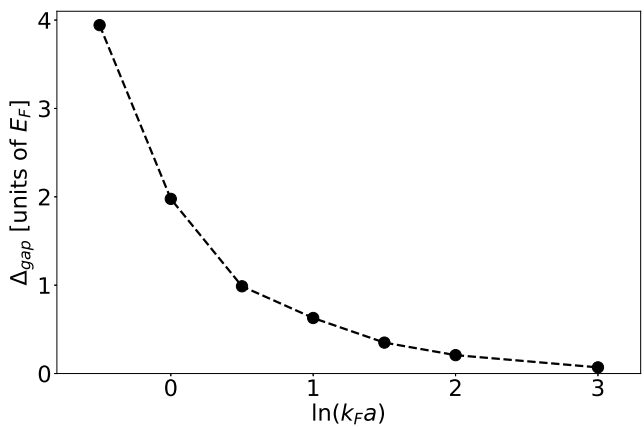}
\caption{DMC pairing gaps throughout the strongly interacting regime.}
\label{fig:gaps_ef}
\end{figure}

To re-cap, it is certainly plausible to use a small $N$ and then
employ a correction on the closed-shell energies as per, say,
Fig.~\ref{fig:finite_err}, i.e., from the non-interacting gas; this would allow one
to make the interacting BCS-side results closer
to the thermodynamic limit. We've already seen in Fig.~\ref{fig:all_E} that the closed-shell interacting results
bear this interpretation out. Something analogous could be done when trying to extract a thermodynamic limit
value for the pairing gap, in which case Fig.~\ref{fig:fg_gaps} would be relevant.
Of course, that would still leave open the question of how to handle finite-size effects away from the deep
BCS regime.
Instead, we have chosen
to approximate the thermodynamic-limit value, throughout the crossover, as the average of our 
best two sets of points, namely $N=50$ and $N=58$ and their neighbors. Beyond simply following
from Fig.~\ref{fig:fg_gaps}, this is an attempt to estimate finite-size effects more generally.
Our overall DMC thermodynamic-limit for the pairing gap is shown 
in Fig.~\ref{fig:gaps_ef}. We can see that the energy of the gap sharply increases as the BEC regime is approached and correspondingly a much reduced pairing gap in the BCS regime. 

\begin{figure}[t]
\includegraphics[width=\columnwidth]{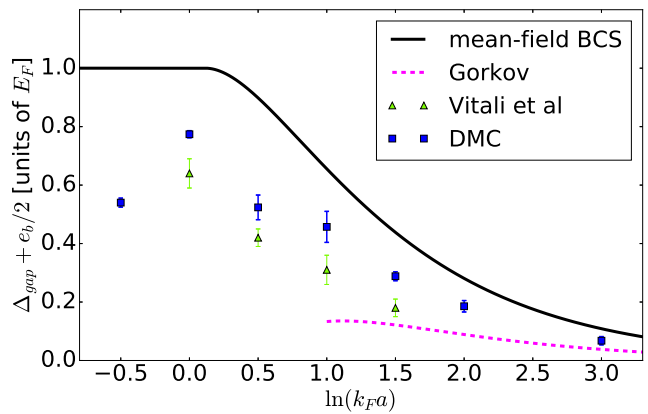}
\caption[Updated pairing gaps compared to literature]{Re-scaled DMC pairing gaps (squares)
compared with the result of Eq.~(\ref{eq:Randeria}) (solid curve), as well as the values of 
Vitali \textit{et al} \cite{Vitali:2017} (triangles) and the prediction of the theory of Gorkov and Melik-Barkhudarov~\cite{Gorkov:1961,Petrov:2003} (dotted curve).}
\label{fig:find_58_50_gap}
\end{figure}

Another choice of dependent variable, which is commonly made in the literature, is to show
$\Delta_{gap} + e_b/2$ vs the interaction strength; this is the same plot
as in Fig.~\ref{fig:meanField} and effectively produces a ``zoomed-in'' version of what's going on. 
In Fig.~\ref{fig:find_58_50_gap} 
we compare our new results to the mean-field BCS value from Eq.~(\ref{eq:Randeria}). Overall,
we find a consistent suppression with respect to the mean-field BCS result. Crucially, this is different 
from our pairing gap values in Ref.~\cite{Galea:2016} (or those of Ref.~\cite{Bertaina:2011}), which tended to become larger than the mean-field
value on the BCS side. In Ref.~\cite{Galea:2016}, our $N=26$ DMC pairing gaps were explicitly described as ``finite-size
uncorrected''
(the same could be said about the pioneering calculations of Ref.~\cite{Bertaina:2011}): 
in the present work we carried out considerably more DMC calculations to approximate
the thermodynamic limit better. Our consistent suppression is similar to what one finds 
using the theory of Gorkov and Melik-Barkhudarov~\cite{Gorkov:1961,Petrov:2003}: in two
dimensions this gives
suppression by a factor of $e$ with respect to the mean-field BCS result. 
We also compare with the values produced in a recent work which employs
imaginary-time Green's functions and analytic continuation~\cite{Vitali:2017}. 
For the intermediate region, where these results were produced,
the two sets of points are qualitatively similar, though the agreement is not perfect.

\section{Conclusion}
\label{sec:QMC}

In summary, we have discussed pairing in two dimensions employing a number of formalisms.
We started from mean-field BCS theory, which was solved (for the first time) for a finite interaction
range; this confirmed that the effective ranges employed in the rest of this work were
sufficiently short so as not to impact the final extraction of pairing gaps. 
We then proceeded to discuss finite-size effects in the strongly interacting crossover, and their
impact on the extraction of the pairing gap via odd-even staggering. More specifically,
we used DMC to determine the ground-state energy at various interaction strengths within the BEC-BCS crossover. We then determined the energy pairing gap for closed-shell values. 
As part of this process, we carried out an investigation of the effect of finite simulation sizes at various interaction strengths. We saw that the finite-size effects closer to the BCS regime closely followed the trend predicted by the non-interacting gas model, and deep into the BEC regime as the attractive interaction became increasingly strong-finite size effects decreased in relative magnitude and followed less closely the trend of the non-interacting gas. The main result is that as we go toward the BCS regime we find a suppression
with respect to the mean-field BCS pairing gap prediction, qualitatively similar to that found 
using the theory of Gorkov and Melik-Barkhudarov.

\begin{acknowledgments}
The authors would like to thank F. Diakonos, S. Giorgini, 
G. Palkanoglou, E. Vitali, and S. Zhang for several discussions. 
This work was supported in part by the Natural Sciences and Engineering Research Council (NSERC) of Canada, the Canada Foundation for Innovation (CFI), and the Early
Researcher Award (ERA) program of the Ontario Ministry of Research, Innovation and Science. Computational resources were provided by SHARCNET and NERSC.
\end{acknowledgments}


\begin{thebibliography}{99}

\bibitem{Bloch:2008}
I.\ Bloch, J.\ Dalibard, and W.\ Zwerger,
Rev.\ Mod.\ Phys.\ {\bf 80}, 885 (2008).

\bibitem{Giorgini:2008}
S.\ Giorgini, L.\ P.\ Pitaevskii, and S.\ Stringari,
Rev.\ Mod.\ Phys.\ {\bf 80}, 1215 (2008).

\bibitem{Levinsen:2014}
J.\ Levinsen and M.\ M.\ Parish,
Annu.\ Rev.\ Cold At.\ Mol.\ {\bf 3}, 1 (2015).

\bibitem{Carlson:2003}
J.\ Carlson, S.\-Y.\ Chang, V.\ R.\ Pandharipande, and K.\ E.\ Schmidt,
Phys.\ Rev.\ Lett.\ {\bf 91}, 050401 (2003).

\bibitem{Chang:2004}
S.\ Y.\ Chang, V.\ R.\ Pandharipande, J.\ Carlson, and K.\ E.\ Schmidt,
Phys.\ Rev.\ A {\bf 70}, 043602 (2004).

\bibitem{Astrakharchik:2004}
G.\ E.\ Astrakharchik, J.\ Boronat, J.\ Casulleras, and and S.\ Giorgini,
Phys.\ Rev.\ Lett. {\bf 93}, 200404 (2004).

\bibitem{Forbes:2011}
M.\ M.\ Forbes, S.\ Gandolfi, and A.\ Gezerlis,
Phys.\ Rev.\ Lett.\ {\bf 106}, 235303 (2011).

\bibitem{Gandolfi:2011}
S.\ Gandolfi, K.\ E.\ Schmidt, and J.\ Carlson,
Phys.\ Rev.\ A {\bf 83}, 041601 (2011).

\bibitem{Forbes:2012}
M.\ M.\ Forbes, S.\ Gandolfi, and A.\ Gezerlis,
Phys.\ Rev.\ A {\bf 86}, 053603 (2012).


\bibitem{Carlson:2012}
J.\ Carlson, S.\ Gandolfi, and A.\ Gezerlis, 
Prog.\ Theor.\ Exp.\ Phys.\ 01A209 (2012).


\bibitem{Pilati:2014}
S.\ Pilati, I.\ Zintchenko, and M.\ Troyer,
Phys.\ Rev.\ Lett.\ {\bf 112}, 015301 (2014).

\bibitem{Schonenberg2:2017}
L.\ M.\ Schonenberg, G.\ J.\ Conduit,
Phys.\ Rev.\ A {\bf 95}, 013633 (2017).

\bibitem{Dawkins:2017}
W. Dawkins and A. Gezerlis, Phys. Rev. A \textbf{96}, 043619 (2017).

\bibitem{Gunter:2005}
K.\ G\"unter, T.\ St\"oferle, H.\ Moritz, M.\ K\"ohl, and T.\ Esslinger,
Phys.\ Rev.\ Lett.\ {\bf 95}, 230401 (2005).

\bibitem{Martiyanov:2010}
K.\ Martiyanov, V.\ Makhalov, and A.\ Turlapov,
Phys.\ Rev.\ Lett.\ {\bf 105}, 030404 (2010).

\bibitem{Frohlich:2011}
B.\ Fr\"ohlich, M.\ Feld, E.\ Vogt, M.\ Koschorreck, W.\ Zwerger, and M.\ K\"ohl,
Phys.\ Rev.\ Lett.\ {\bf 106}, 105301 (2011).

\bibitem{Feld:2011}
M.\ Feld, B.\ Fr\"ohlich, E.\ Vogt, M.\ Koschorreck, and M.\ K\"ohl,
Nature.\ {\bf 480}, 75-78 (2011).

\bibitem{Orel:2011}
A.\ A.\ Orel, P.\ Dyke, M.\ Delehaye, C.\ J.\ Vale, and H.\ Hu,
New J.\ Phys.\ {\bf 13}, 113032 (2011).

\bibitem{Dyke:2011}
P.\ Dyke, E.\ D.\ Kuhnle, S.\ Whitlock, H.\ Hu, M.\ Mark, S.\ Hoinka, M.\ Lingham, P.\ Hannaford, and C.\ J.\ Vale,
Phys.\ Rev.\ Lett.\ {\bf 106}, 105304 (2011).

\bibitem{Sommer:2012}
A.\ T.\ Sommer, L.\ W.\ Cheuk, M.\ J.\ H.\ Ku, W.\ S.\ Bakr, and M.\ W.\ Zwierlein,
Phys.\ Rev.\ Lett.\ {\bf 108}, 045302 (2012).

\bibitem{Makhalov:2014}
V.\ Makhalov, K.\ Martiyanov, and A.\ Turlapov,
Phys.\ Rev.\ Lett.\ {\bf 112}, 045301 (2014).

\bibitem{Boettcher:2015}
I.\ Boettcher, L.\ Bayha, D.\ Kedar, P.\ A.\ Murthy, M.\ Neidig, M.\ G.\ Ries, A.\ N.\ Wenz, G.\ Z\"urn, S.\ Jochim, and T.\ Enss,
Phys.\ Rev.\ Lett.\ {\bf 116}, 045303 (2016).

\bibitem{Ong:2015}
W.\ Ong, C.-Y.\ Cheng, I.\ Arakelyan, and J.\ E.\ Thomas,
Phys.\ Rev.\ Lett.\ {\bf 114}, 110403 (2015).

\bibitem{Murthy:2015}
P.\ A.\ Murthy, I.\ Boettcher, L.\ Bayha, M.\ Holzmann, D.\ Kedar, M.\ Neidig, M.\ G.\ Ries, A.\ N.\ Wenz, G.\ Z\"urn, and S.\ Jochim,
Phys.\ Rev.\ Lett.\ {\bf 115}, 010401 (2015).

\bibitem{Ries:2015}
M.\ G.\ Ries, A.\ N.\ Wenz, G.\ Z\"urn, L.\ Bayha, I.\ Boettcher, D.\ Kedar, P.\ A.\ Murthy, M.\ Neidig, T.\ Lompe, and S.\ Jochim,
Phys.\ Rev.\ Lett.\ {\bf 114}, 230401 (2015).

\bibitem{Fenech:2016} K. Fenech, P. Dyke, T. Peppler, M. G. Lingham, S. Hoinka, H. Hu, and C. J. Vale, Phys. Rev. Lett. \textbf{116}, 045302 (2016).

\bibitem{Boettcher:2016} I. Boettcher, L. Bayha, D. Kedar, P. A. Murthy, M. Neidig, M. G. Ries, A. N. Wenz, G. Z\"urn, S. Jochim, and T. Enss, Phys. Rev. Lett. \textbf{116}, 045303 (2016).


\bibitem{Martiyanov:2016} K. Martiyanov, T. Barmashova, V. Makhalov, and A. Turlapov, Phys. Rev. A \textbf{93}, 063622 (2016).

\bibitem{Cheng:2016} C. Cheng, J. Kangara, I. Arakelyan, and J.~E. Thomas, Phys. Rev. A \textbf{94}, 031606 (2016). 


\bibitem{Luciuk:2017}
C.\ Luciuk, S.\ Smale, F.\ Böttcher, H.\ Sharum, B.\ A.\ Olsen, S.\ Trotzky, T.\ Enss, and J.\ H.\ Thywissen,
Phys.\ Rev.\ Lett.\ {\bf 118}, 130405 (2017).

\bibitem{Toniolo:2017}
U.\ Toniolo, B.\ C.\ Mulkerin, C.\ J.\ Vale, X.-J.\ Liu, H.\ Hu,
Phys.\ Rev.\ A {\bf 96}, 041604(R) (2017).

\bibitem{Nishida:2008}
Y.\ Nishida and S.\ Tan,
Phys.\ Rev.\ Lett.\ {\bf 101}, 170401 (2008).

\bibitem{Liu:2010} X.-J. Liu, H. Hu, and P. D. Drummond, Phys. Rev. B \textbf{82}, 054524 (2010).

\bibitem{Valiente:2011} M. Valiente, N. T. Zinner, and K. Molmer, Phys. Rev. A \textbf{84}, 063626 (2011).

\bibitem{Bertaina:2011}
G.\ Bertaina and S.\ Giorgini,
Phys.\ Rev.\ Lett.\ {\bf 106}, 110403 (2011).

\bibitem{Werner:2012}
F.\ Werner and Y.\ Castin,
Phys.\ Rev.\ A\ {\bf 86}, 013626 (2012).


\bibitem{Bauer:2014}
M.\ Bauer, M.\ M.\ Parish, and T.\ Enss,
Phys.\ Rev.\ Lett.\ {\bf 112}, 135302 (2014).

\bibitem{Shi:2015}
H.\ Shi, S.\ Chiesa, and S.\ Zhang,
Phys.\ Rev.\ A {\bf 92}, 033603 (2015).

\bibitem{Anderson:2015}
E.\ R.\ Anderson and J.\ E.\ Drut,
Phys.\ Rev.\ Lett.\ {\bf 115}, 115301 (2015).

\bibitem{He:2015}
L.\ He, H.\ L\"u, G.\ Cao, H.\ Hu, X.-J.\ Liu,
Phys.\ Rev.\ A {\bf 92}, 023620 (2015).

\bibitem{Rammelmuller:2015} L. Rammelm\"uller, W. J. Porter and J. E. Drut, Phys.\ Rev.\ A, \textbf{93}, 033639 (2016).

\bibitem{He:2016} L. He, Ann. Phys. (N.Y.) \textbf{373}, 470 (2016).

\bibitem{Galea:2016}
A. Galea, H. Dawkins, S. Gandolfi, and A. Gezerlis, Phys. Rev. A \textbf{93}, 023602 (2016).

\bibitem{Klawunn:2016}
M.\ Klawunn,
Phys.\ Lett.\ A.\ {\bf 380}, 34 (2016).

\bibitem{Galea:2017}
A. Galea, T. Zielinski, S. Gandolfi, and A. Gezerlis, J Low Temp Phys \textbf{189}, 451 (2017).

\bibitem{Vitali:2017}
E.\ Vitali, H.\ Shi, M.\ Qin, S.\ Zhang, Phys.\ Rev.\ A {\bf 96}, 061601(R) (2017)

\bibitem{Schonenberg:2017}
L.\ M.\ Schonenberg, P.\ C.\ Verpoort, G.\ J.\ Conduit,
Phys.\ Rev.\ A {\bf 96}, 023619 (2017).

\bibitem{Drut:2018}
J.\ E.\ Drut, J.\ R.\ McKenney, W.\ S.\ Daza, C.\ L.\ Lin, C.\ R.\ Ordóñez,
Phys.\ Rev.\ Lett.\ {\bf 120}, 243002 (2018).

\bibitem{jeszenszki:2019}
P.\ Jeszenszki, A.\ Alavi, J.\ Brand,
Phys.\ Rev.\ A {\bf 99}, 033608 (2019).

\bibitem{Gezerlis:2008}
A.\ Gezerlis and J.\ Carlson,
Phys.\ Rev.\ C {\bf 77}, 032801 (2008).

\bibitem{Gandolfi:2015}
 S.\ Gandolfi, A.\ Gezerlis, and J.\ Carlson, 
 Ann.\ Rev.\ Nucl.\ Part.\ Sci.\ \textbf{65}, 303 (2015).

\bibitem{Buraczynski:2016}
M.\ Buraczynski and A.\ Gezerlis,
Rev.\ Lett.\ {\bf 116}, (2016).

\bibitem{Horikoshi:2017}
M. Horikoshi, M. Koashi, H. Tajima, Y. Ohashi, and M. Kuwata-Gonokami,
Phys.\ Rev.\ X {\bf 7}, 041004 (2017).

\bibitem{Miyake:1983}
K.\ Miyake,
Prog.\ Theor.\ Phys.\ {\bf 69}, 1794 (1983).

\bibitem{Randeria:1989}
M.\ Randeria, J.-M.\ Duan, and L.-Y.\ Shieh,
Phys.\ Rev.\ Lett.\ {\bf 62}, 981 (1989);
Phys.\ Rev.\ B {\bf 41}, 327 (1990).

\bibitem{Petrov:2003}
D.\ S.\ Petrov, M.\ A.\ Baranov, and G.\ V.\ Shlyapnikov,
Phys.\ Rev.\ A {\bf 67}, 031601(R) (2003).

\bibitem{Gorkov:1961} L. P. Gorkov and T. K. Melik-Barkhudarov, JETP, {\bf 40}, 1452 (1961) [Soviet Phys. JETP {\bf 13}, 1018 (1961)].



\end{thebibliography}
\end{document}